\documentclass[onecolumn,
 showpacs,amsmath,amssymb,11pt]{revtex4}
\usepackage{graphicx}
\usepackage{epstopdf}
\usepackage{bm}
\usepackage{subfigure}
\usepackage[latin1]{inputenc}
\usepackage{rotating}
\usepackage{longtable}
\usepackage{float}
\usepackage[normalem]{ulem}
\usepackage{dcolumn}
\usepackage{bm}
\usepackage{color}
\usepackage{footnote}

\usepackage{color, soul}

\begin{document}

\preprint{APS/123-QED}

\title{Characterization of Cs vapor cell coated with octadecyltrichlorosilane using coherent population trapping spectroscopy}
\author{Moustafa Abdel Hafiz$^1$, Vincent Maurice$^1$, Ravinder Chutani$^1$, Nicolas Passilly$^1$, Christophe Gorecki$^1$, St\'ephane Gu\'erandel$^2$, Emeric de Clercq$^2$ and Rodolphe Boudot$^1$}

\affiliation{$^1$FEMTO-ST, CNRS, UFC, 26 chemin de l'Epitaphe 25030 Besan\c{c}on Cedex, France \\ $^2$LNE-SYRTE, Observatoire de Paris, CNRS, UPMC, 61 avenue de l'Observatoire, 75014 Paris, France}

\date{\today}

\begin{abstract}
We report the realization and characterization using coherent population trapping (CPT) spectroscopy of an octadecyltrichlorosilane (OTS)-coated centimeter-scale Cs vapor cell. The dual-structure of the resonance lineshape, with presence of a narrow structure line at the top of a Doppler-broadened structure, is clearly observed. The linewidth of the narrow resonance is compared to the linewidth of an evacuated Cs cell and of a buffer gas Cs cell of similar size. The Cs-OTS adsorption energy is measured to be (0.42 $\pm$ 0.03) eV, leading to a clock frequency shift rate of $2.7\times10^{-9}/$K in fractional unit. A hyperfine population lifetime, $T_1$, and a microwave coherence lifetime, $T_2$, of 1.6 and 0.5 ms are reported, corresponding to about 37 and 12 useful bounces, respectively. Atomic-motion induced Ramsey narrowing of dark resonances is observed in Cs-OTS cells by reducing the optical beam diameter. Ramsey CPT fringes are detected using a pulsed CPT interrogation scheme. Potential applications of the Cs-OTS cell to the development of a vapor cell atomic clock are discussed.
\end{abstract}


\maketitle
\clearpage
\section{Introduction}
The detection of long-lived relaxation time spin-polarized atoms in alkali vapor cells is of critical importance in a relevant number of fundamental physics experiments and applications including optical magnetometry \cite{Budker:Nature:2007, Kominis:Nature:2003}, vapor-cell atomic frequency standards \cite{VanierAudoin}, quantum information storage \cite{Julsgaard:Nature:2001}, quantum teleportation \cite{Krauter:Nature:2013}, slow-light applications \cite{Budker:PRL:1999, Xiao:PRL:2008}, investigation of atomic parity non conservation \cite{Bouchiat:PL:1982} or spin-squeezing \cite{Kuzmich:PRL:2000}. In alkali vapor cells, two main techniques are used to slow down the fast wall-induced depolarization of the atoms. The first method is to dilute the alkali vapor in a buffer gas. In the so-called Dicke regime \cite{Dicke:PR:1953}, slight and frequent alkali-buffer gas collisions result in a slow diffusive motion of atoms in the cells and contribute to increase the time for alkali atoms to collide against the cell walls.  Nevertheless, the presence of a buffer gas induces a frequency shift and broadening of the optical resonance signal \cite{Pitz:PRA:2009} that degrades the optical pumping efficiency and consequently the stability of atomic clocks and magnetometers but also a temperature-dependent frequency shift of the atomic clock hyperfine transition frequency \cite{Kozlova:PRA:2011, Beverini:OC:1981, Vanier:JAP:1982}. These drawbacks are particularly critical in chip-scale atomic devices \cite{Knappe:MEMS:2007} where high buffer gas pressures (up to 700 Torr) are required in order to minimize the wall-induced relaxation. The second method consists in coating the glass walls of the resonance cell with a chemically-inert anti-relaxation film material in order to make the atoms experience a significant number of bounces before complete destruction of the observed atomic population or coherence. Paraffin-coatings such as polyethylene, Paraflint [CH$_3$(CH$_2$)$_n$CH$_3$] or tetracontane [CH$_3$(CH$_2$)$_{38}$CH$_3$], formed by long-chain alkane molecules, were pioneerly proposed by Ramsey \cite{Ramsey:RSI:1950}, first demonstrated by Robinson et al.\cite{Robinson:APS:1958}, studied by Brewer \cite{Brewer:JCP:1963} and later extensively by Bouchiat et Brossel \cite{Bouchiat:PR:1966}. These coatings have demonstrated to support up to 10$^4$ atom-wall collisions, leading to observed linewidths in centimeter-scale cells of a few Hertz on either Zeeman \cite{Budker:RMP:2002} or hyperfine transitions \cite{Robinson:APL:1982}. More recently, Balabas et al. reported exceptional anti-relaxation properties of alkene-based coatings (C$_n$H$_{2n+1}$) demonstrating polarized alkali-metal vapor with minute-long transverse Zeeman population and coherence lifetimes in a 3 cm diameter cell, which corresponds to about 10$^6$ polarization preserving bounces \cite{Balabas:PRL:2010, Balabas:OE:2010}. For clock applications, it has to be noted that comparable microwave hyperfine frequency shift and linewidths have been recently measured between paraffin-coated and alkene-coated Rb cells \cite{Corsini:PRA:2013, Budker:PRA:2005}.
However, a drawback of paraffin and alkene-based coatings, recently studied and investigated in detail using surface science techniques by a large scientific community \cite{Seltzer:JCP:2010}, is their relatively low-temperature melting point (about 85$^{\circ}$C and 35-100$^{\circ}$C  \cite{Balabas:PRL:2010, Balabas:OE:2010} respectively). This feature prevents them from being used in microfabricated vapor cells because of the high temperature of the anodic bonding
process \cite{Dziuban}. In addition, cell temperatures higher than 80$^{\circ}$C are usually required in miniature atomic clocks. For this purpose, octadecyltrichlorosilanes (OTS) layers, thermally stable up to 170$^{\circ}$C in presence of Rb vapor \cite{Seltzer:JAP:2009}, are interesting candidates. Seltzer et al. reported the measurement of up to 2100 collisions before the population relaxes ($T_1$) in a K vapor cell with OTS multilayers \cite{Seltzer:PRA:2007} and typically 25 bounces with OTS monolayers in Rb vapor \cite{Seltzer:JAP:2008}. In the presence of Rb, multilayer and monolayer OTS were found to withstand 170$^{\circ}$C and 190$^{\circ}$C, respectively \cite{Seltzer:JAP:2009}. Through measurements of hyperfine resonance linewidths and frequency shifts, Yi et al. demonstrated in a 8 mm side cubic Rb vapor cell that Rb atoms collide up to 40 times with the cell walls before coherence relaxation and estimated their adsorption energy to be 0.065 eV \cite{Yi:JAP:2008}. From relaxation rate measurements in Rb vapor cells, Camparo et al. reported about 5 collisions with other silane materials \cite{Camparo:JAP:1987}. More recently, a microfabricated vapor cell with OTS anti-relaxation coating was reported with the demonstration of 11 surviving wall collisions \cite{Straessle:APL:2014} in a double resonance Rb clock setup.\\
In FEMTO-ST, we proposed an original buffer-gas filled Cs vapor microcell fabrication and filling technology \cite{Hasegawa:SA:2011} that uses post-activation of Cs vapor in the hermetically-sealed cell through local laser heating of a Cs pill dispenser. Since no alkali vapor is present during the anodic bonding process, we can expect from the conclusions of \cite{Fedchak} that the OTS coating should withstand the temperature of 350$^{\circ}$C. This procedure could be well-adapted for the development of OTS-coated microcells since anodic bonding process could be operated in optimal conditions at such elevated temperatures. In that sense, we investigate here the use of OTS coatings in Cs vapor cells for compact CPT-based atomic clocks applications and potentially later for miniature atomic clocks. Surprisingly, we found only one article in the literature where interactions between OTS coating and Cs atoms were studied \cite{Stephens:JAP:1994}.\\
This article aims to report the detection, spectroscopy and hyperfine clock frequency shift measurements of coherent population trapping (CPT) resonances in centimeter-scale OTS-coated Cs vapor cells. Section \ref{sec:setup} describes the cell coating and filling procedure and the experimental set-up used to perform CPT spectroscopy. Section \ref{sec:res} reports using CPT spectroscopy the characterization of a Cs-OTS cell through measurements of the CPT resonance linewidth, adsorption energy of the coating, $T_1$ and $T_2$ relaxation times. The detection of Ramsey fringes in the Cs-OTS cell is reported. The potential of the Cs-OTS cell in atomic clocks applications is finally discussed.

\section{Experimental set-up}
\label{sec:setup}
\subsection{Wall coating and filling of Cs vapor cells}
\begin{figure}[h!]
\centering
\includegraphics[width=\linewidth]{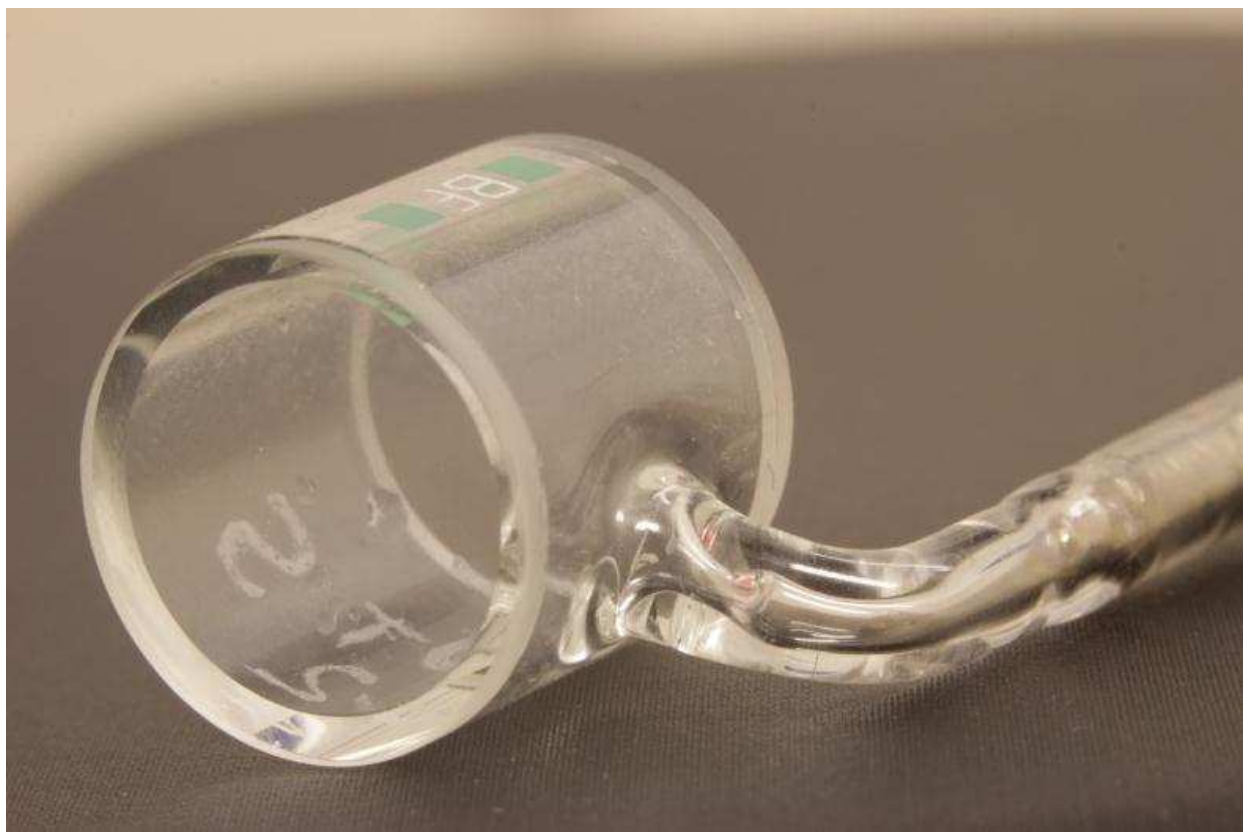}
\caption{(Color online) Photograph of the OTS-coated cell.}
\label{fig:Photocell}
\end{figure}

We produced and tested 3 different cylindrical vapor cells made of borosilicate glass (Borofloat\circledR33, Schott). The first cell is a reference evacuated Cs cell with a diameter of 15 mm and a length of 15 mm. The second cell, named Cs-OTS, with the same dimensions than the evacuated Cs cell, is equipped with OTS anti-relaxation coating and does not contain any buffer gas. The last cell, named Cs-N$_2$-Ar, is a Cs vapor cell filled with a N$_2$-Ar buffer gas mixture of total pressure 15 Torr and pressure ratio $r$ = P$_{Ar} /$ P$_{N_2}$ = 0.4. Its dimensions are a diameter and a length of 10 mm. The coating and filling procedure for the Cs-OTS cell was performed as follows. First, the cell was cleaned with piranha solution (H$_2$SO$_4$/H$_2$O$_2$) and rinsed with de-ionized water. The cleaned cell was dried in an oven at 150$^{\circ}$C for 20 min. The  cell was hermetically connected to a bottle containing a few microliters of OTS (from Sigma-Aldrich Inc.). This step was performed in a glovebox under a dry nitrogen (N$_2$) atmosphere to prevent OTS oxidation. Still connected to the bottle, the cell was baked in an oven at 150$^{\circ}$C for 4 h to let OTS evaporate and polymerize on the cell inner surface. The cell was later returned to the glovebox, disconnected from the bottle, rinsed with acetone and dried. Since the inner diameter of the capillary linking the cell to its sidearm is narrow (typically 600 $\mu$m), the introduction of any liquid inside the cell is impeded by capillary action. Consequently, the sidearm of the cell was immersed in the solution and the pressure in the glovebox was alternatively decreased and increased to allow the gas contained in the cell to be gradually replaced by the solution. Once coated, the cell was connected to a glass manifold evacuated with a turbomolecular pump below 10$^{-7}$ mbar. Cesium (from Alfa Aesar Inc.) was distilled in the sidearms of the cells before being finally sealed off. Figure \ref{fig:Photocell} shows a photograph of the OTS-coated cell. The presence of the coating material is confirmed by the existence of "milky" patches, layers or thin droplets on the cell inner walls.

\subsection{Experimental apparatus}
\begin{figure}[h!]
\centering
\includegraphics[width=\linewidth]{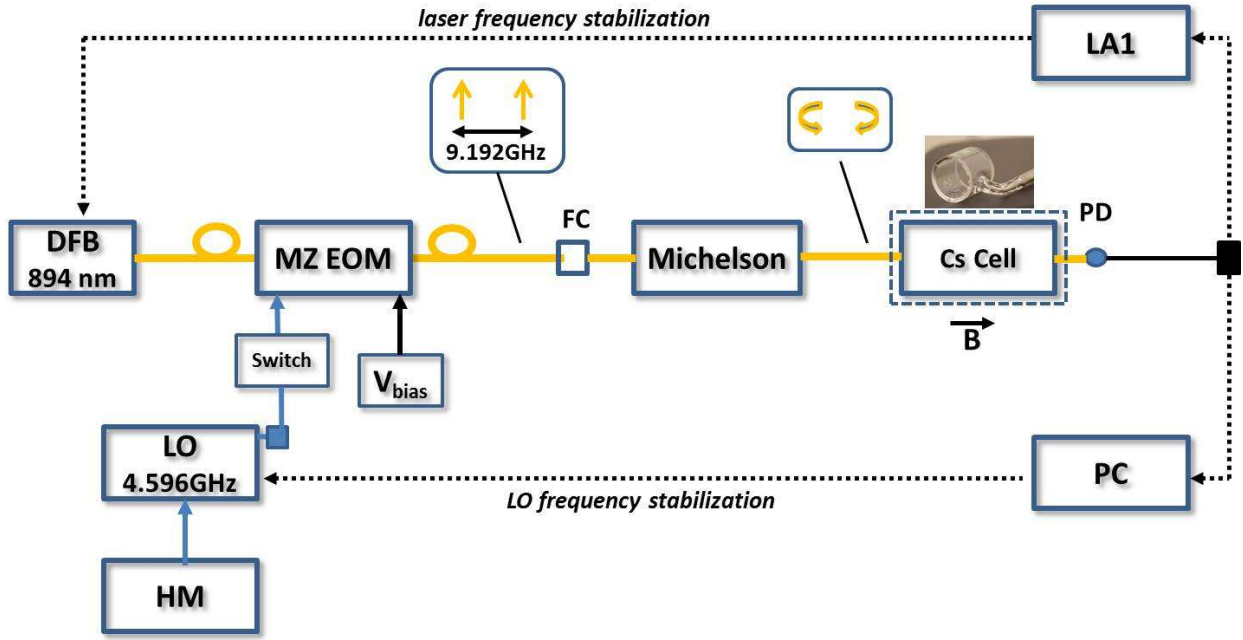}
\caption{(Color online) Experimental setup used to detect CPT resonances in Cs vapor cells. DFB: Distributed-Feedback diode laser, MZ EOM: Mach-Zehnder electro-optic modulator, Michelson: Michelson delay-line and polarization orthogonalizer system, $V_{bias}$: dc bias voltage applied onto the EOM to reject the optical carrier, LO: local oscillator, FC: fiber collimator, PD: photodiode, LA1: lock-in amplifier, PC: personal computer. A microwave switch can be used to turn on and off the 4.596 GHz signal in order to make the atoms interact with a sequence of optical pulse trains.}
\label{fig:cpt-setup}
\end{figure}

Figure \ref{fig:cpt-setup} shows the experimental set-up used to perform coherent population trapping spectroscopy or relaxation time measurements in the Cs vapor cells. The laser source is a distributed-feedback (DFB) diode laser tuned on the Cs D$_1$ line at 894.6 nm \cite{Liu:IM:2012}. The D$_1$ line of alkali atoms is known to be a better candidate than the D$_2$ line for CPT interaction \cite{Stahler:OL:2002}. An optical isolator, not shown in Fig. \ref{fig:cpt-setup}, is placed at the output of the diode laser to prevent optical feedback. The laser beam is sent into a pigtailed Mach-Zehnder intensity electro-optic modulator (MZ EOM - Photline NIR-MX800-LN-10). The latter is driven at 4.596 GHz by a low noise microwave frequency synthesizer, referenced to a high-performance hydrogen maser available in the laboratory, in order to generate two first-order optical sidebands frequency-split by 9.192 GHz for CPT interaction. A direct digital synthesis (DDS) allows the fine tuning of the output signal frequency. The optical carrier is actively rejected with an extinction ratio of 24 dB using the lock-in microwave detection technique described in Ref. \cite{Liu:PRA:2013}. At the output of the EOM, light is sent through a Michelson delay-line and polarization orthogonalizer system to produce the so-called push-pull optical pumping interaction scheme \cite{Jau:PRL:2004}. This technique allows to increase greatly the number of atoms participating in the 0-0 magnetic field insensitive clock transition. The diameter of the output beam is expanded to 1.5 cm using a telescope to cover the whole cell diameter. A variable neutral density filter is added to produce laser power variations and a diaphragm is inserted in order to change the laser beam diameter. The laser beam is sent into the Cs vapor cell to ensure atom-light interaction. The cell is implemented into an oven temperature-stabilized at the millikelvin level with a high-precision temperature controller inspired from Ref. \cite{Boudot:RSI:2005}. A homogeneous longitudinal static magnetic field is applied using a solenoid to lift the degeneracy of the ground-state Zeeman manifold. A magnetic field flux density value of 1.55 G (for both the Cs cell and the Cs-OTS cell) and 100 mG (for the buffer-gas cell) was applied in order to separate and discriminate correctly Doppler-broadened ground-state microwave Zeeman transitions. The assembly is surrounded by a double-layer mu-metal magnetic shield to protect atoms from environmental magnetic perturbations. The light transmitted through the cell is detected by a photodiode. The laser is frequency stabilized on the bottom of the Doppler-broadened optical line using a lock-in amplifier based (LA1) synchronous modulation-demodulation technique that corrects the laser bias current. CPT resonance spectroscopy is realized by scanning slowly the 4.596 GHz output frequency using the computer-controlled DDS. The synthesizer output microwave signal can be stabilized to the CPT clock resonance by correcting permanently the DDS frequency. Inspired by Ref. \cite{Liu:OE:2013}, the 4.596 GHz microwave signal power can be switched on and off and the MZ EOM used as a light switch in order to make the atoms interact with a sequence of optical pulse trains. This technique is convenient to perform relaxation times measurements through the Franzen's technique of relaxation in the dark \cite{Franzen} or to produce a Ramsey-like pulsed CPT interrogation.

\section{Experimental results}
\label{sec:res}
\subsection{Continuous regime CPT spectroscopy}
\begin{figure}[h!]
\centering
\includegraphics[width=\linewidth]{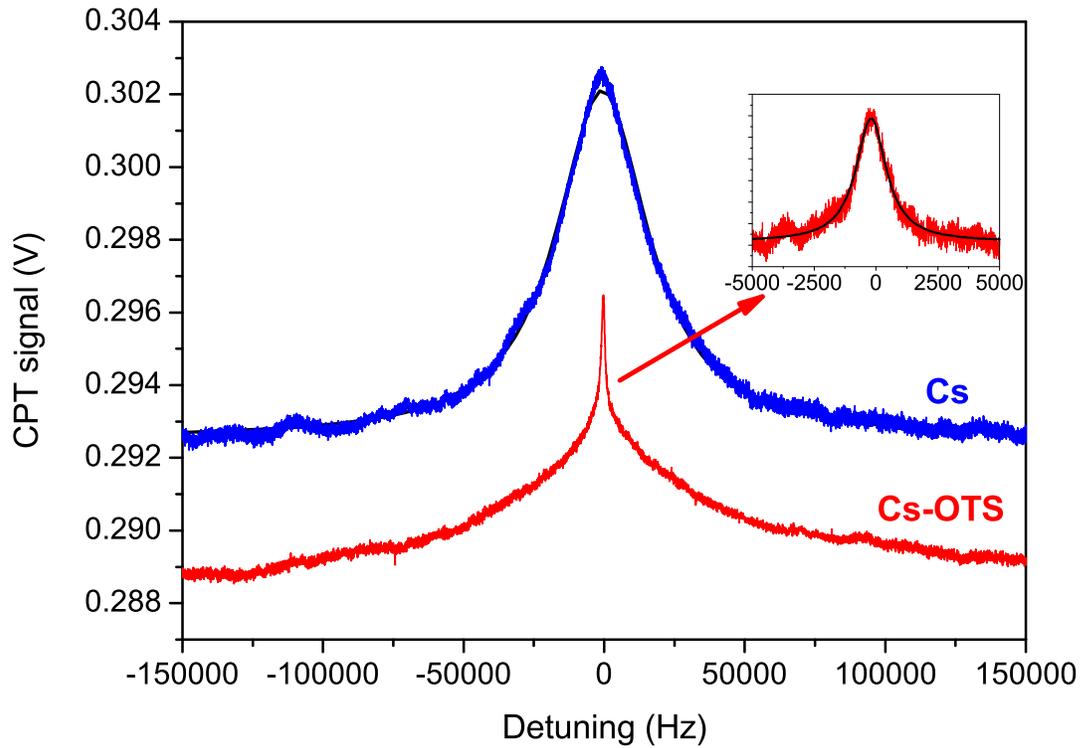}
\caption{(Color online) CPT clock resonance in the Cs-OTS cell compared to the one detected in the evacuated Cs cell. The laser power incident on the cell is 100 $\mu$W. The cell temperature is 35$^{\circ}$C. Experimental data for the evacuated Cs cell are approximated by a Lorentzian fit function. The narrow structure of the resonance in the Cs-OTS cell (figure inset) is approximated by a Lorentzian fit function. The laser beam diameter is 15 mm. For the Cs-OTS cell, the slight asymetry of the CPT resonance is attributed to a minor thermal transient of the clock set-up table during the scan of the CPT resonance.}
\label{fig:im1}
\end{figure}

\begin{figure}[h!]
\centering\includegraphics[width=\linewidth]{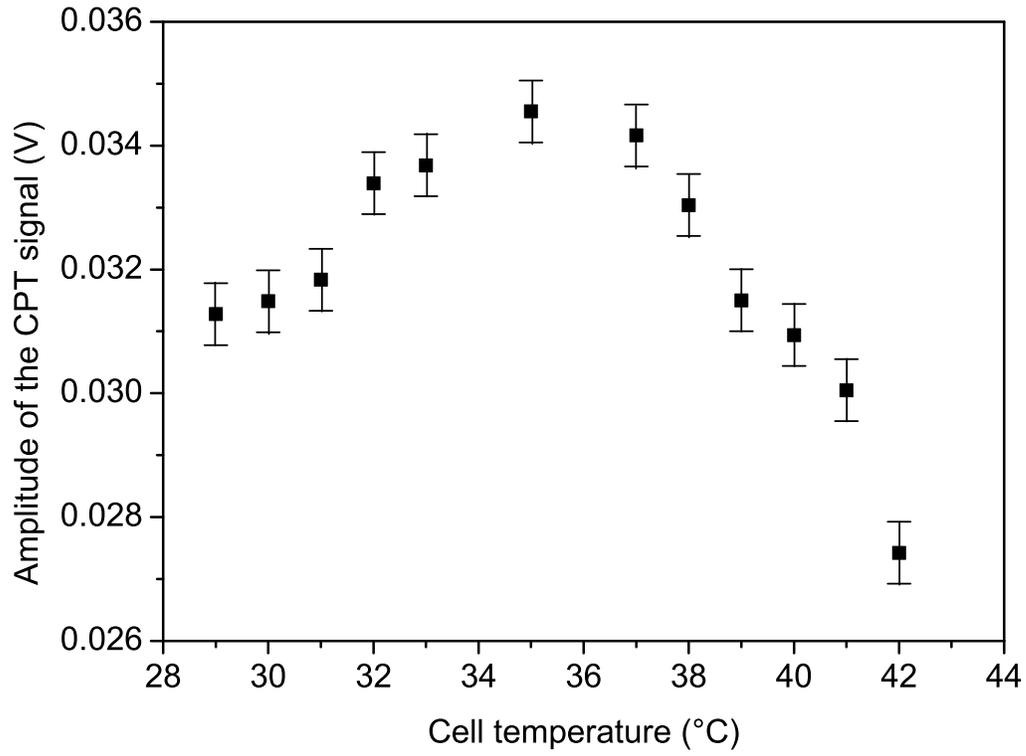}
\caption{Amplitude of the CPT signal (narrow structure) in the Cs-OTS cell versus the cell temperature. The laser power is 600 $\mu$W. The static magnetic field is 892 mG.}
\label{fig:ots-signal-T}
\end{figure}

\begin{figure}[h!]
\centering
\includegraphics[width=\linewidth]{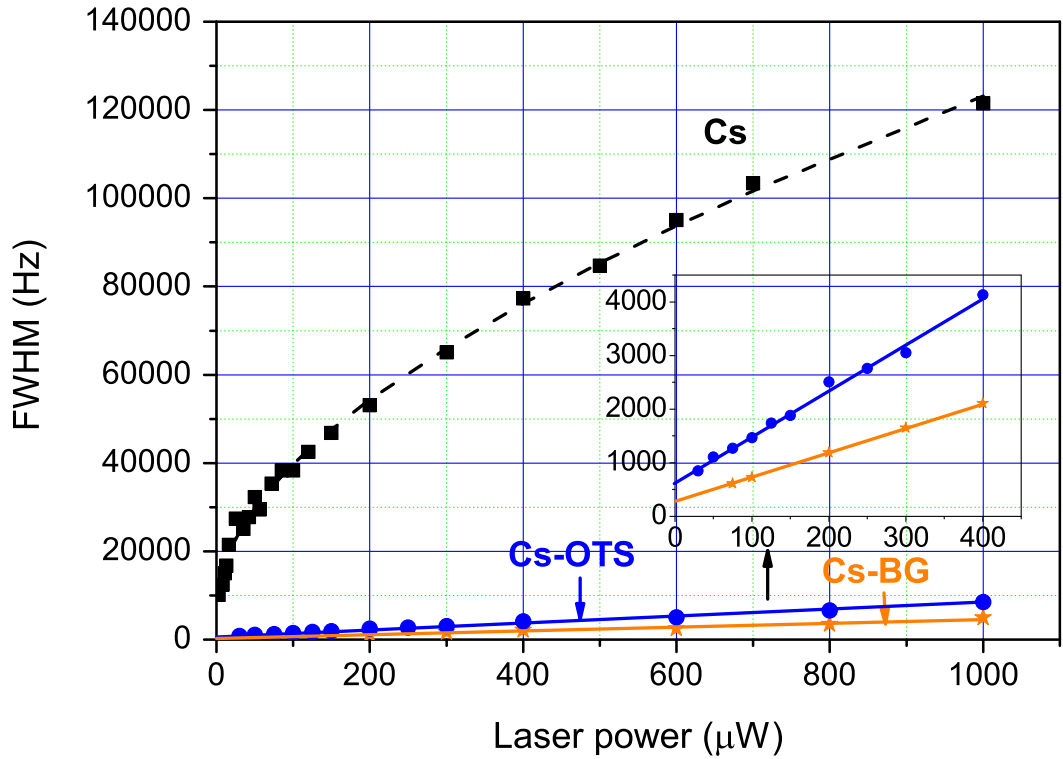}
\caption{(Color online) Linewidth of the CPT resonance in vapor cells versus the laser power for: evacuated Cs cell (squares), Cs-OTS cell narrow structure (circles), Cs-N$_2$-Ar cell (stars). Error bars are covered by data points. The cell temperature is 35$^{\circ}$C. Data of the evacuated Cs cell are fitted by the function $a+b\sqrt{P}+c P$ (dashed line, $P$ the laser power, $a$, $b$, and $c$ are constants). Experimental data in the Cs-OTS cell and the buffer-gas filled cell are fitted by a linear function. The figure inset is a zoom for Cs-OTS and Cs-N$_2$-Ar cells.}
\label{fig:im3}
\end{figure}

Figure \ref{fig:im1} reports, for identical experimental conditions, the CPT clock resonance in the Cs-OTS cell and in the evacuated Cs cell for a total incident laser power of 100 $\mu$W. Both cells exhibit about the same off-resonance background level, i.e. the same laser power absorption. The CPT resonance in the pure Cs cell is well approximated by a Lorentzian function. The dual-structure of the dark resonance, signature of the anti-relaxation effect, is obvious in the OTS-coated cell. The pedestal of the resonance is characterized by a Doppler-broadened structure whereas the top of the resonance is narrowed thanks to the coating material. As shown in the figure inset, the narrow structure is well approximated by a Lorentzian function with a linewidth of 1486 Hz.\\
Figure \ref{fig:ots-signal-T} shows the signal of the CPT resonance (narrow structure) in the Cs-OTS cell versus the cell temperature. The amplitude of the CPT signal is found to be maximized for a cell temperature of about 35$^{\circ}$C.  The same behavior was observed for the broad structure signal. The contrast of the narrow structure was found to increase from 1.4 to 2.6 \% from 29 to 42$^{\circ}$C. Note that the CPT linewidth of both narrow and broad structures were measured to be reduced very slightly with temperature in the 30 - 50$^{\circ}$C range. The CPT signal-to-linewidth ratio is optimized at 35$^{\circ}$C, and most of the following results are obtained at this temperature.\\
For further investigation, Figure \ref{fig:im3} shows the linewidth of the CPT resonance in the three different Cs vapor cells versus the total laser power $P$ for a cell temperature of 35$^{\circ}$C. In an evacuated Cs cell, the main effects contributing to the CPT resonance linewidth are the Doppler broadening, the power broadening, the atom-light limited transit time and the Cs-Cs spin-exchange collisions. In a Doppler-broadened system, as explained in Ref. \cite{Javan:PRA:2002} for a similar experiment of electromagnetically induced transparency (EIT) and in \cite{Radonjic:PRA:2009}, it can be shown that the CPT linewidth is proportional to the square root of intensity for low laser intensity of the driving field and is independent of the Doppler width. This is an effect similar to the laser-induced line narrowing effect \cite{Kazakov:JPB:2007}. At higher laser intensities, the usual power broadening effect is recovered with a CPT linewidth proportional to the laser intensity. This behavior is clearly demonstrated in our measurement where the CPT linewidth - laser power dependence curve shows two distinct regimes with a sudden linewidth decrease for laser powers lower than about 50 $\mu$W. In this laser power region, the contribution of the Doppler broadening is gradually reduced as the laser intensity is decreased. For a cell temperature of 35$^{\circ}$C, the  CPT linewidth extrapolated at null laser power in the evacuated Cs cell is measured to be 7 kHz, in excellent agreement with theoretical calculations yielding a total linewidth of 7.07 kHz (sum of spin-exchange and transit-time contributions). The transit time contribution is given by $1/(\pi t_t)$ where $t_t$ is the mean time of flight between two wall collisions. We note $t_t= \ell/v_m=45 \, \mu$s with $v_m = \sqrt{8kT/(\pi m)} = 222$ m/s being the mean atomic velocity, $T$ the cell temperature, $k$ the Boltzmann constant and $m$ the mass of the Cs atom. We note $\ell =10$ mm the mean free path between two wall collisions in a cylindrical cell of radius $R$ and length $L$ such as $1/\ell = (1/R+1/L)/2$ \cite{VanierAudoin}. The transit time limited linewidth is 7.05 kHz. The spin exchange relaxation term is calculated at 35$^{\circ}$C following \cite{Vanier:PRA:1998} as:
\begin{equation}
\Delta \nu_{se}=  (11/16)n_{Cs}v_r \sigma_{se}/\pi.
\label{se}
\end{equation}
Here $n_{Cs}$ is the Cs density ($1.2 \times 10^{11}$ atom/cm$^3$ \cite{Taylor:PR:1937}), $v_r$ is the average relative velocity of Cs atoms, $v_r=\sqrt{2} v_m=313$ m/s, and $\sigma_{se}$ is the Cs spin-exchange cross-section ($2.18\times 10^{-14}$ cm$^2$ \cite{VanierAudoin}), yielding $\Delta \nu_{se}= 18$ Hz. For the Cs-OTS cell and the buffer-gas filled Cs cell, experimental widths in Hz are well fitted by linear functions 634 + 8.6 $P$ and 268.9 + 4.6 $P$, respectively, with $P$ in $\mu$W. The CPT linewidth extrapolated at null laser power is about 10 times narrower in the Cs-OTS cell compared to the pure Cs cell but 2.4 wider than in the buffer-gas filled Cs cell. In the Cs-OTS cell, the laser power broadening is about 13 times smaller than in the evacuated Cs cell and 2 times bigger than in the Cs-N$_2$-Ar cell.\\

In a cell coated with an anti-relaxation material, atoms collide with the surface, stick to it for a characteristic time $\tau_s$ and eventually return to the vapor. Attraction of the atom to the cell walls is mainly dominated by the long-range van der Walls force $F$ that depends on the dielectric constant $\varepsilon$ of the wall and the electric dipole operator $D$ of the atom such as $F \propto (\varepsilon-1)/(\varepsilon+1) D^2$ \cite{Stephens:JAP:1994}. The Cs energy levels are shifted during the atom-surface interaction, so that the phase of the hyperfine coherence of  the escaping atom is shifted by a mean amount $\Phi$ at each collision, which depends on the surface characteristics. Reducing the wall surface attraction through the choice of a surface with low polarizability and choosing a coating material with low dielectric constant will help to reduce the atom-wall interaction time and to make the collision more elastic. This physisorption process can be characterized by an adsorption energy $E_a$, related to the sticking time $\tau_s$, that traduces the kinetic energy an atom must have to escape the coating surface attraction and in turn the period the alkali atoms spend physically adsorbed on the wall of the cell. The adsorption energy can be estimated from the measurement of the clock frequency shift $\delta \nu$, defined as the difference between the actual clock frequency and the exact unperturbed Cs atom frequency (9.192631770GHz), versus the cell temperature $T$ as \cite{Goldenberg:PR:1961,Vanier:PRA:1974}:
\begin{equation}
| \delta \nu | \propto E_a \exp [E_a / kT]
\label{Ea}
\end{equation}

\begin{figure}[h!]
\centering
\includegraphics[width=\linewidth]{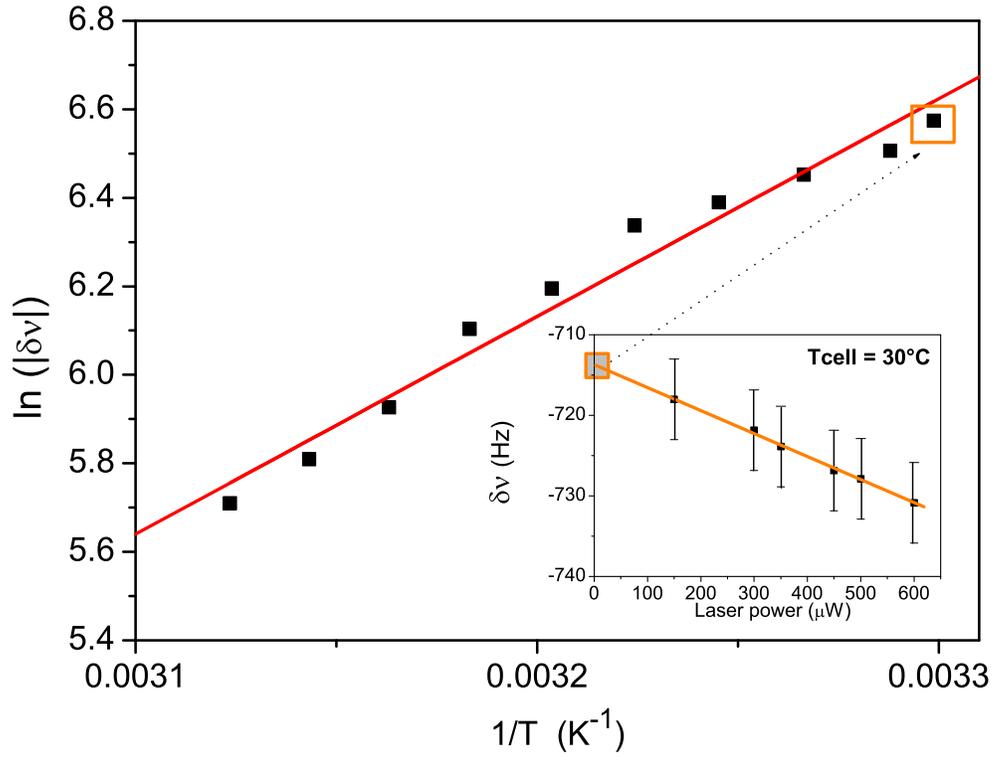}
\caption{(Color online) Logarithm of the frequency shift $|\delta \nu|$ versus the inverse of the cell temperature $1/T$ with $T$ in Kelvins. Experimental data are fitted by a linear function from which is extracted the value of $E_a$. In the main figure, error bars are covered by data points. The inset shows an example for a cell temperature of 30$^{\circ}$C of light-shift measurement used to extract the temperature frequency shift at null laser power.}
\label{fig:Tshift-OTS1}
\end{figure}

\begin{figure}[h!]
\centering
\includegraphics[width=\linewidth]{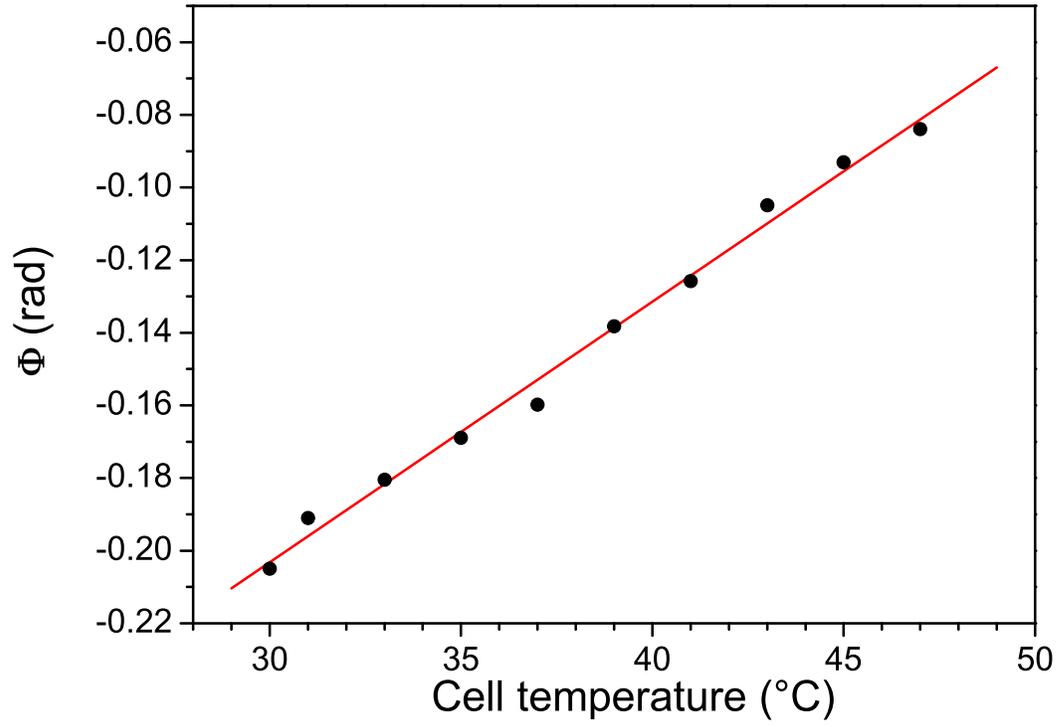}
\caption{(Color online) Phase shift per wall collision as a function of the cell temperature. The solid line is a linear fit. Error bars are covered by data points. }
\label{fig:phase-T}
\end{figure}

We measured the clock frequency versus the cell temperature (from 30 to 47$^{\circ}$C) to extract the value of $E_a$. The temperature of the Cs reservoir was kept slightly lower than the temperature of the cell wall to prevent deposition of Cs onto the wall coating. For each cell temperature, the clock frequency is measured for different values of laser intensity and extrapolation to null laser intensity is performed as shown on the inset of the figure \ref{fig:Tshift-OTS1}. A negative frequency shift of a few hundreds of Hertz is measured as usually observed on wall-coated cells \cite{Corsini:PRA:2013, Yi:JAP:2008}. At high temperatures, an atom spends less time on the wall and is found to experience a smaller frequency shift magnitude. The shift rate is measured to be $+24.6$ Hz/K. According to Eq. (\ref{Ea}), the plot of $\ln(|\delta \nu|)$ against $1/T$ (see Fig. \ref{fig:Tshift-OTS1}) is a straight line of slope $E_a/k$. The fitting data of Fig. \ref{fig:Tshift-OTS1} leads to $E_a=0.42$ eV with a statistical uncertainty of 0.03 eV. This value corresponds to a sticking time of the atom on the surface $\tau_s = \tau_0 \exp{\frac{E_a}{kT}} \sim$ 29 $\mu$s, where $\tau_0 \sim 10^{-12}$ s \cite{Stephens:JAP:1994}. The value of $E_a$ is in good agreement with the value (0.40 $\pm$ 0.03 eV) reported by Stephens et al. \cite{Stephens:JAP:1994}, the only one we found in the literature for Cs-Pyrex-OTS. It is worth to note that the measurement of \cite{Stephens:JAP:1994} is based on a completely independent method, namely the atomic number density measurement. For Cs-Paraflint interaction, $E_a$ was estimated to be 0.8 eV \cite{Goldenberg:PR:1961}. This value remains higher than the typical value of about 0.1 eV reported by other researchers in Rb vapor cells coated with paraffin \cite{Bouchiat:PR:1966, Brewer:JCP:1963} or even 0.065 eV in OTS-coated Rb cells \cite{Yi:JAP:2008}.\\
The frequency shift of the clock transition is related to the mean phase shift per wall collision $\Phi$ as \cite{Goldenberg:PR:1961,Vanier:PRA:1974,Budker:PRA:2005}:
\begin{equation}
\delta \nu = \Phi/(2 \pi t_t).
\label{phase}
\end{equation}
The phase shift values, computed from data of Fig. \ref{fig:Tshift-OTS1}, are shown in Fig. \ref{fig:phase-T} as a function of the cell temperature. On the limited experimental temperature range around $39^\circ$C the phase shift can be fitted by a straight line,
\begin{equation}
\Phi = -0.138(2)+0.0074(3)(T_{cell}-39),
\label{phase}
\end{equation}
with $T_{cell}$ the cell temperature in degree Celsius.  $\Phi = (- 168 \pm 2)$ mrad/collision at $35^\circ$C and  $(- 94\pm 2)$ mrad/collision at $45^{\circ}$C. For comparison, $\Phi$ was measured to be $(90\pm 10)$ mrad/collision for Cs-Paraflint coating \cite{Goldenberg:PR:1961} at an unspecified temperature, $-$19 mrad/collision \cite{Straessle:APL:2014} and $-$ 65 mrad/collision \cite{Yi:JAP:2008} in Rb-OTS cells of smaller dimensions heated at 60$^{\circ}$C. $\Phi$ is proportional to $E_a \exp [E_a / kT]$ and $E_a$ is proportional to the polarizability of the atom \cite{Goldenberg:PR:1961}. As cesium has the largest polarizability a larger phase shift is expected for cesium than for rubidium or potassium.\\
The actual phase shift experienced by an atom colliding the wall (adiabatic collision) is assumed to follow a Gaussian distribution of mean $\Phi$ and of variance $\Phi^2$ \cite{Goldenberg:PR:1961,Budker:PRA:2005}. Here we do not take into account the atomic velocity distribution and $t_t$ is considered as a constant.
After $n$ collisions the accumulated phase shift distribution $G(n,\varphi)$ is also Gaussian of mean $n \Phi$ and of variance $n \Phi^2$. The probability of experiencing  $n$ collisions during a time $t$ is assumed to obey a Poisson distribution $p(n,t)=exp(-t/t_t)(t/t_t)^n/n!$  \cite{Budker:PRA:2005}. As a result, the mean  and the variance of the phase shift after a time $t$ are:
\begin{eqnarray}
\begin{split}
& \varphi_m =\Sigma_{n=0}^{\infty}p(n,t)n\Phi= (t/t_t)\Phi,\\
& \sigma_{\varphi}^2 =<\varphi^2>-\varphi_m^2=2(t/t_t)\Phi^2.
\end{split}
\label{phasem}
\end{eqnarray}
The lineshape $L_S $ of the CPT resonance is given by the weighted sum of detuned Lorentzian profiles  $L(\delta, \gamma)$:
\begin{equation}
L_S(\delta, \gamma)=\Sigma_{n=0}^{\infty}\left( p(n,t) \int_{-\infty}^{+\infty} L(\delta-\varphi/t_t, \gamma)G(n,\varphi)d\varphi \right) ,
\label{Lorz}
\end{equation}
with $\delta$ the angular frequency Raman detuning, and $\gamma$ the half-width of the resonance. We have no analytical expression for $L_S$, which is a weighted sum of Voigt profiles. Nevertheless, as shown in Fig. \ref{fig:im1} the lineshape is well fitted by a Lorentzian profile. Budker \textit{et al.} \cite{Budker:PRA:2005} have shown that the contribution of the phase dispersion to the line width of a "classical" microwave transition is (in Hz) \cite{Rahman:QE:1987,Budker:PRA:2005}:
\begin{equation}
\Delta \nu_{\Phi} =\frac{ \Phi^2}{\pi t_t}.
\label{gammaphase}
\end{equation}
At null laser intensity the width of a CPT resonance or a "classical" resonance are given by the same coherence relaxation terms, thereby we can assume that Eq. (\ref{gammaphase}) is valid in our case. The full resonance width (FWHM) $\Delta \nu$ can be written as the sum of different contributions:
\begin{equation}
\Delta \nu =\Delta \nu_{w}+\Delta \nu_{\Phi}+\Delta \nu_{se}+\Delta \nu_{st}.
\label{fullwidth}
\end{equation}
Here, these terms are evaluated for a cell temperature of $35^{\circ}$C. $\Delta \nu_{st}$ is the relaxation term due to atoms incoming in the stem. We assume that an atom impacting the stem region, of radius $r$, is lost and that this probability is proportional to the stem fractional area  with respect to the whole cell area. It follows $\Delta \nu_{st}=\frac{1}{\pi t_t}\frac{r^2}{2R(R+L)}$. We estimate $r\simeq 0.8$ mm and $\Delta \nu_{st}=13$ Hz. The spin-exchange term $\Delta \nu_{se}$ is about 18 Hz. The dephasing  term of Eq. (\ref{gammaphase}) is 200 Hz. The $\Delta \nu_{w}$ term takes into account other relaxation contributions induced by wall collisions, \textit{e. g.} population and hyperfine coherence relaxations or velocity changes such that the atoms go out of optical resonance by Doppler effect and are lost for the signal. As $\Delta \nu= 634$ Hz at null laser intensity, $\Delta \nu_{w}$ is estimated to be about 400 Hz. The coating can be characterized by a mean number $n$ of useful bounces before the loss of the atom or the loss of atomic phase memory.  $n$ is equal to the ratio of the relaxation term due to a single wall collision to the relaxation term due to wall collisions in the coated cell, $n\approx \frac{1/ t_t}{\pi(\Delta \nu_{w}+\Delta \nu_{\Phi})}$. We get $n\approx12$ bounces. This result is of the same order of magnitude as the one (11 bounces) reported by Straessle et al. in Rb-OTS cells \cite{Straessle:APL:2014} and as those (20-30 bounces) reported in Ref. \cite{Yi:JAP:2008}, and much smaller than results obtained with paraffin or alkenes.

\subsection{Measurements of population lifetime $T_1$}
\label{sT1T2}
\begin{figure}[h!]
\centering
\includegraphics[width=\linewidth]{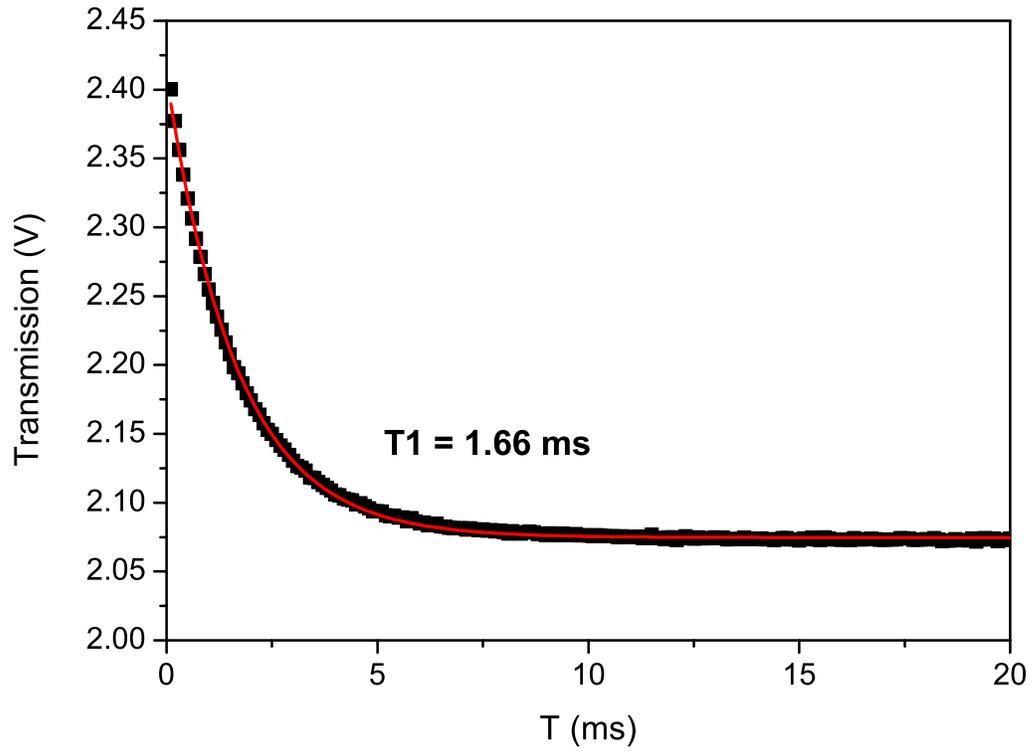}
\caption{(Color online) Measurement of the $T_1$ relaxation time in the Cs-OTS cell. The total laser power incident on the cell is 500 $\mu$W. The actual power seen by the atoms  (power of one sideband) is about one half of the total laser power. The cell temperature is 35$^{\circ}$C. The static magnetic field is 900 mG. Squares: data, solid line: fitted exponential function. }
\label{fig:T1}
\end{figure}

\begin{figure}[h!]
\centering
\includegraphics[width=\linewidth]{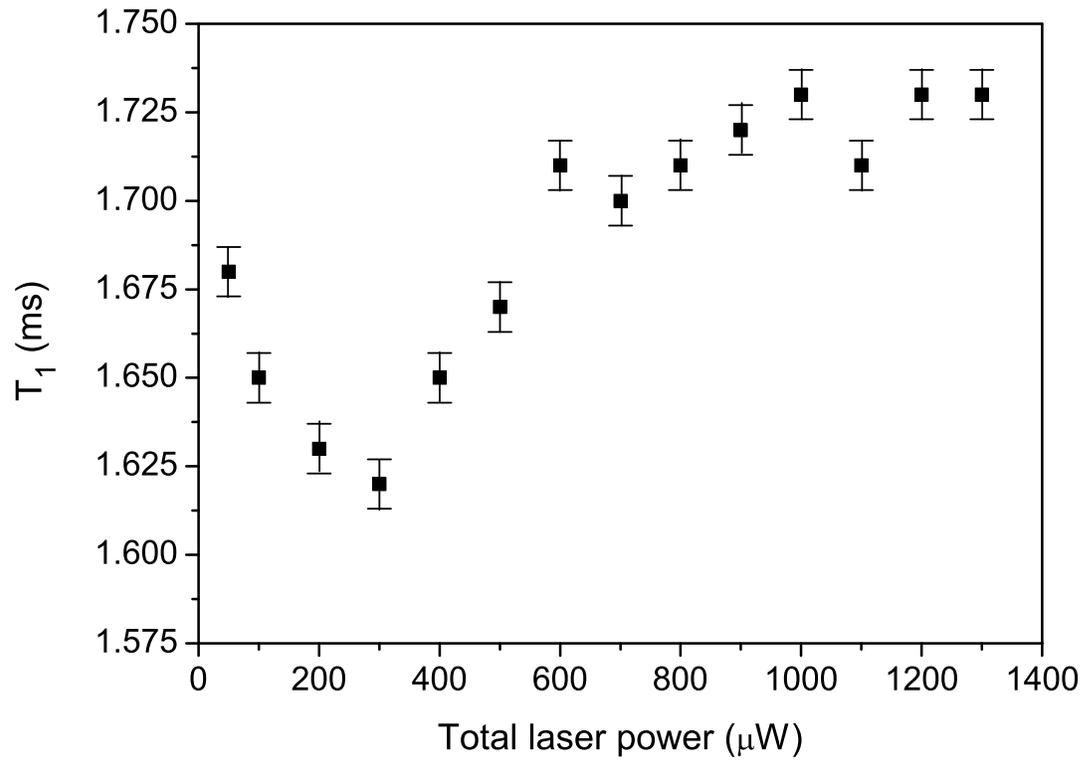}
\caption{ Measurement of the $T_1$ relaxation time in the Cs-OTS cell versus the total laser power incident on the cell. The actual power seen by the atoms is about one half of the total laser power. The cell temperature is 35$^{\circ}$C. The static magnetic field is 900 mG.}
\label{fig:T1-P}
\end{figure}

Additional characterization of the OTS-coated cell was realized through measurements of $T_1$ relaxation time in the dark using the Franzen's technique. Atoms interact with a sequence of optical pulse trains. Atoms are first optically pumped during a constant $\tau_p$ duration pulse ($\tau_p$ = 3 ms) in the hyperfine ground state ($F = 4$) with a single laser frequency. Then, light is switched off and atoms relax (($F = 4$) population) in the dark during a dark time $T$. Each cycle, the duration of the time $T$ is slightly incremented. The next light pulse allows to detect the atomic signal by measuring the laser power transmitted through the cell. The signal is measured 20 $\mu$s after the beginning of the pulse. It is defined as the average value of 25 successive measurements realized in a 25 $\mu$s duration measurement window of the same light pulse. This study allows to measure how the atomic system, initially prepared in a determined initial state, evolves to the Boltzmann equilibrium. It is a measure of the hyperfine population relaxation performed on all Zeeman sublevels.\\
Figure \ref{fig:T1} reports, for an incident laser power of 500 $\mu$W, the measurement of the relaxation time $T_1$ in the dark of the hyperfine level ($F = 4$)
population. A single laser frequency is obtained by detuning the carrier frequency of 9 GHz, tuning consequently a single sideband to the atomic resonance. For information, we noted that the polarization scheme (circularly polarized beam or push-pull optical pumping) and the static magnetic value had a negligible impact on the measured value of the $T_1$ relaxation time. Experimental data are well fitted by an exponential decay function with a time constant $T_1$ = (1.66 $\pm$ 0.007) ms. Such a time constant corresponds to about 37 useful bounces. Figure \ref{fig:T1-P} shows the evolution of the measured $T_1$ value function of the laser power. For laser power higher than 300 $\mu$W, the value of $T_1$ is measured to increase slightly with the laser power before saturation above 1 mW. On the opposite, below 300 $\mu$W, the $T_1$ value increases for decreasing laser power. Clearly, there are two opposite phenomena that are not yet identified and need further investigations.

\subsection{Motion-induced Ramsey narrowing and Ramsey spectroscopy in Cs-OTS cells}
\begin{figure}[h!]
\centering
\subfigure[]{\includegraphics[width=0.9\linewidth]{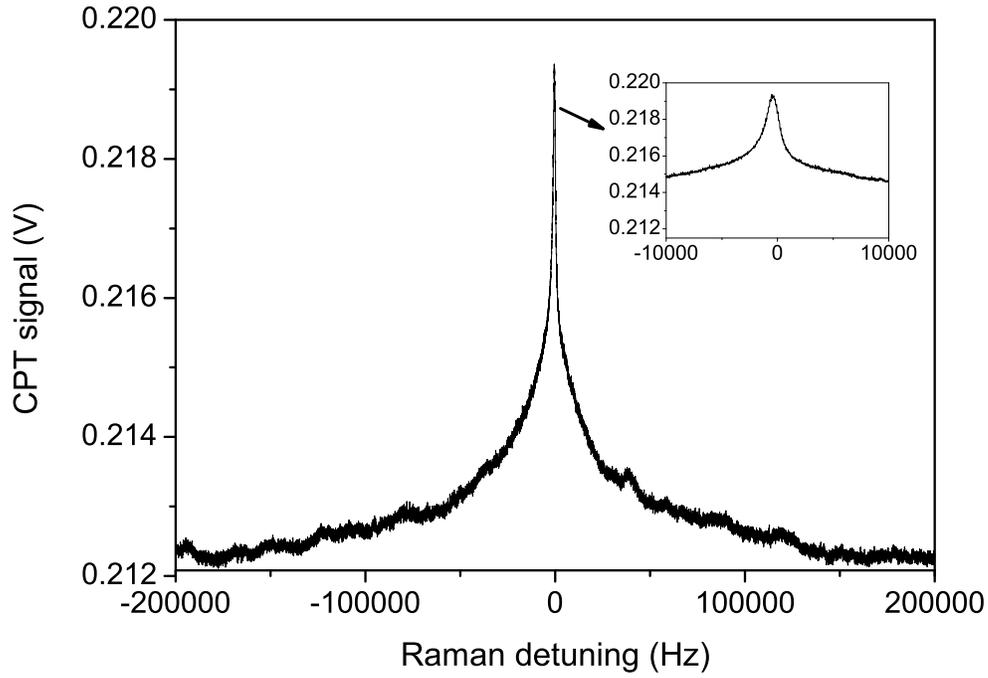}
\label{fig:im7}} \vfill
\subfigure[]{\includegraphics[width=0.9\linewidth]{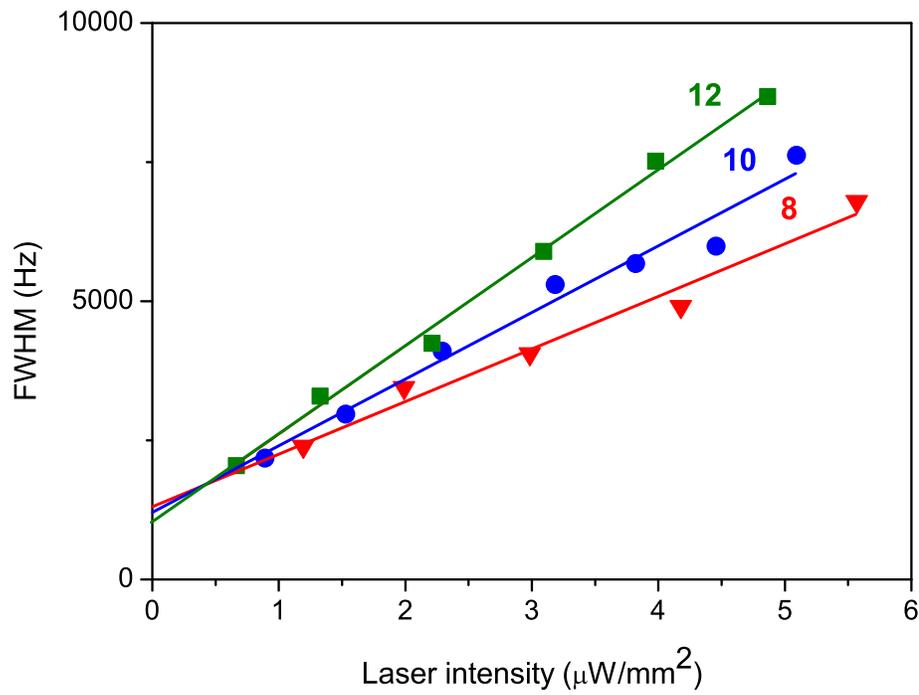}
\label{fig:im8}}
 \caption{(Color online) (a): CPT resonance in the Cs-OTS cell for a beam diameter of 8 mm. The laser power is 100 $\mu$W. (b) Linewidth of the CPT resonance narrow structure versus the laser intensity for different laser beam diameters (8, 10 and 12 mm). Error bars are covered by data points.}
\end{figure}

\begin{figure}[h!]
\centering
\includegraphics[width=\linewidth]{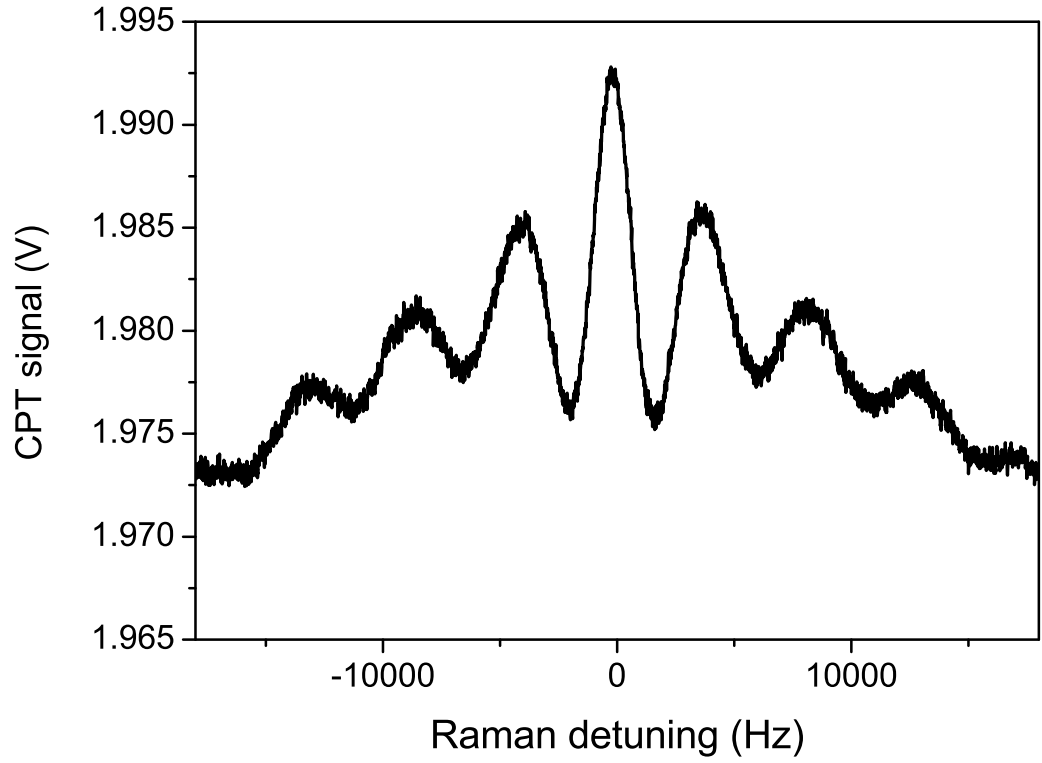}
\caption{ CPT-Ramsey fringe detected in a Cs-OTS cell. $\tau_p$ = 2 ms, $T_R$ = 0.2 ms. The laser power is 650 $\mu$W. The cell temperature is 35$^{\circ}$C. The beam diameter covers the whole cell diameter.
}
\label{fig:5}
\end{figure}

\begin{figure}[h!]
\centering
\includegraphics[width=\linewidth]{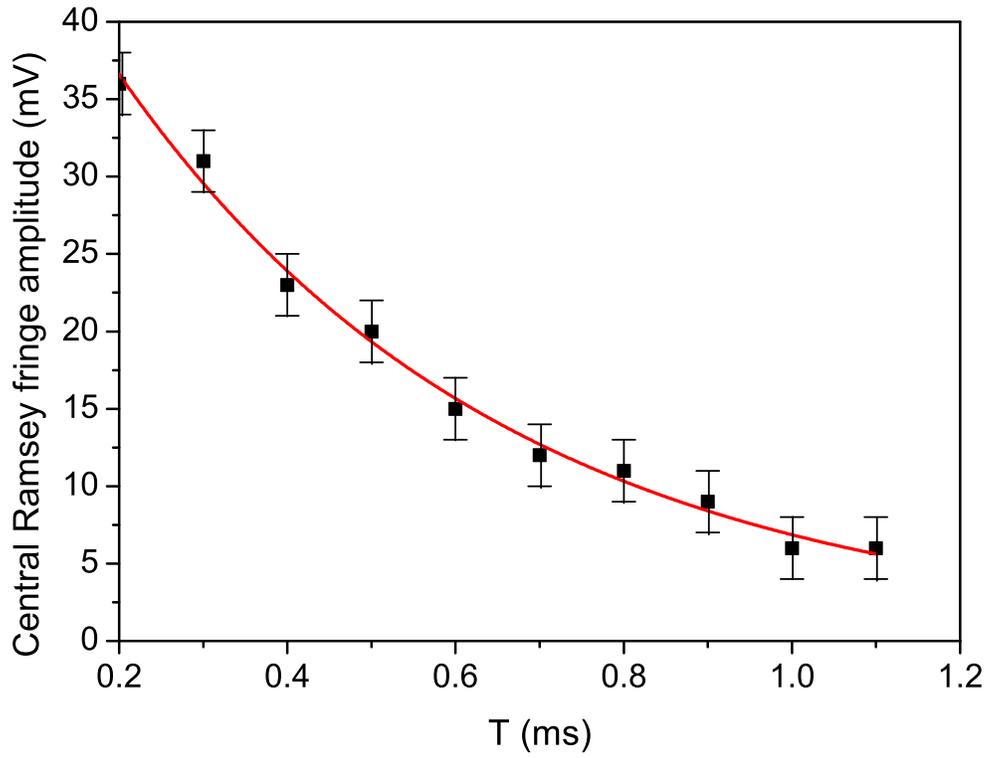}
\caption{(Color online) Amplitude of the central CPT-Ramsey fringe detected in a Cs-OTS cell versus the Ramsey time
$T_R$. The laser power is 1.4 mW. Experimental data are fitted by an exponential decay function of time constant $T_2$ = 0.46 ms. }
\label{fig:T2-ots}
\end{figure}

In a wall-coated cell without buffer gas, the ballistic transport mechanism is distinct from the diffusive behavior in buffer gas-filled cells. Atoms move randomly from wall to wall with constant velocity and direction. For small beam diameters compared to the cell diameter, repeated interactions of atoms with the light fields are equivalent to a sequence of randomly spaced Ramsey pulses in which the two optical fields are turned on and off. The free motion of the polarized atomic spins in and out of the optical interaction region before spin relaxation induces a well-known Ramsey narrowing of the CPT resonance, well studied and described in paraffin-coated cells \cite{Breschi:PRA:2010, Klein:PRA:2011}. Additionally, it has to be noted that in our experiments, the residual Doppler effect on the clock transition is reduced by a Dicke-type narrowing because the cell characteristic length (1.5 cm) is smaller than the Cs atom ground-state microwave wavelength (3.2 cm).\\
We investigated in Cs-OTS cells the CPT resonance lineshape for different beam diameters. Figure \ref{fig:im7} shows a CPT resonance in the Cs-OTS cell for a beam diameter of 8 mm and a laser power of 100 $\mu$W. The lineshape has a Doppler-broadened structure of linewidth 43 kHz, in correct agreement with the atom transit time through the interaction volume. The narrow central peak exhibits a width of 1924 Hz. Figure \ref{fig:im8} shows the linewidth of the CPT resonance narrow structure versus the laser intensity for different laser beam diameters (8, 10 and 12 mm). For low laser intensities, which is the case for each experimental data point of the figure, the width of the narrow structure increases linearly with laser intensity. Most significantly, we observe a decrease of the linewidth with the volume of the interaction region at constant optical intensity. Indeed, smaller beams allow longer phase evolution of atoms in the dark, making the CPT narrow structure lineshape narrower. Simultaneously, as observed in \cite{Breschi:PRA:2010}, we measured an increase of the CPT resonance amplitude with the laser beam radius. This can be explained qualitatively as an increase of the number of interacting atoms in the interrogation volume. On the contrary, we observed that the broad structure linewidth is increased with smaller beam diameter, which corresponds to a reduced atom-light interaction time. The Ramsey picture of bright and dark times also indicates that the narrow structure linewidth should saturate with increasing intensity, rather than continue power broadening as it would in most vapor-cell systems. This occurs because, for sufficient intensity, light fields optically pump atoms into the dark state in a single pass through the beam. This behavior was clearly observed in Ref. \cite{Klein:PRA:2011}.\\
Additionally, the experimental set-up described above was used to detect CPT-Ramsey fringes in the Cs-OTS cell. Figure \ref{fig:5} shows a CPT-Ramsey fringe detected in the Cs-OTS cell. Atoms interact with light in a pulsed CPT sequence where the CPT pumping time $\tau_p$ is 3 ms and the free evolution time in the dark
$T_R$ is 0.2 ms. The laser power is 650 $\mu$W. Figure \ref{fig:T2-ots} shows the evolution of the central Ramsey fringe amplitude versus the Ramsey time $T_R$. Experimental data are well fitted by an exponential decay function with a time constant of (0.46 $\pm$ 0.05) ms. This value can be interpreted as an experimental estimation of the CPT hyperfine coherence lifetime $T_2$. This value of $T_2$ is in correct agreement and slightly smaller than the one extracted from the zero-intensity CPT linewidth measurement of (634 $\pm$ 50) Hz reported in Fig. \ref{fig:im3}, which yields a $T_2$ time of $T_2 = 1 / (\pi \Delta \nu)$ = (0.50 $\pm$ 0.04) ms.
Note that the contribution of the phase shift dispersion term to the relaxation rate of the Ramsey fringes is the same as in the CW interrogation case. When the phase distribution is Gaussian of variance $\sigma_{\varphi}^2$, it can be shown that the fringe amplitude scales as $e^{-\sigma_{\varphi}^2/2}$. Here $\sigma_{\varphi}^2/2 =(t/t_t)\Phi^2$, see (\ref{phasem}), which is equivalent to a time constant $t_t/\Phi^2$ , \textit{i.e.} the same as the one of
the continuous measurement deduced from (\ref{gammaphase}).

\subsection{Applications to atomic clocks}
\begin{figure}[h!]
\centering
\subfigure[]{\includegraphics[width=0.57\linewidth]{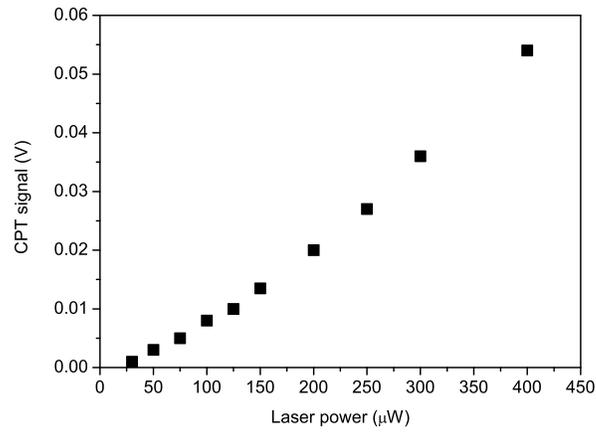}
\label{fig:Signal-OTS}} \vfill
\subfigure[]{\includegraphics[width=0.57\linewidth]{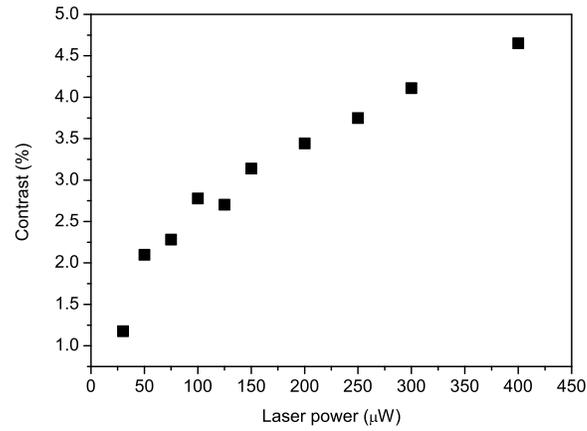}
\label{fig:C-OTS}} \vfill
\subfigure[]{\includegraphics[width=0.57\linewidth]{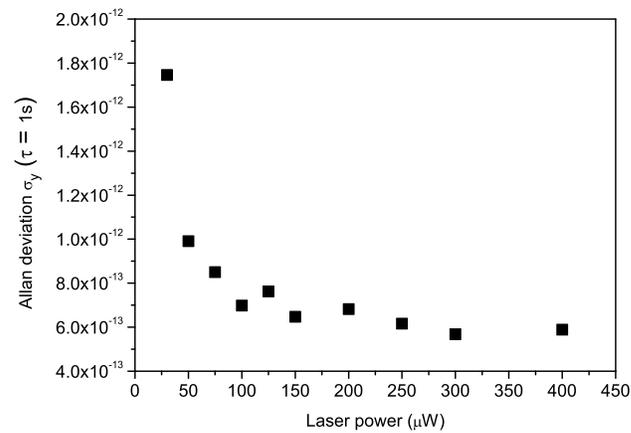}
\label{fig:Allan-OTS}}
 \caption{(a): CPT resonance signal, (b) CPT resonance contrast, (c) calculated shot-noise limited clock frequency Allan deviation versus the laser power incident on the cell. The cell temperature is 35$^{\circ}$C. The laser beam diameter equals the cell diameter. Error bars are covered by data points.}
\end{figure}

Figures \ref{fig:Signal-OTS} and \ref{fig:C-OTS} show the signal and the contrast (defined as the ratio between the CPT signal amplitude and the off-resonance dc background) respectively of the CPT resonance in CW regime for different values of the laser power in the Cs-OTS cell, when the beam diameter covers the whole cell diameter. The signal amplitude shows a quadratic dependence for laser power below 50 $\mu$W while the dependence is linear above this value. This behavior is in good agreement with the CPT signal expression derived in Ref. \cite{Vanier:PRA:1998}. As observed in buffer-gas filled Cs vapor cells using push-pull optical pumping technique, the resonance contrast is measured to increase with the laser power while a saturation value is expected to be observed at higher intensities \cite{Liu:PRA:2013}.\\
The short-term relative frequency stability of an atomic clock, in terms of Allan deviation $\sigma_y (\tau)$, is given by:
\begin{equation}
\sigma_y (\tau) \approx \frac{\Delta \nu}{\nu_0} \frac{1}{SNR} \tau^{-1/2}
\end{equation}
where $\nu_0$ is the clock transition frequency (about 9.192 GHz for Cs atom), $\Delta \nu$ is the resonance full-width at half maximum, $SNR$ is the signal-to-noise ratio of the detected resonance in a 1 Hz bandwidth and $\tau$ is the averaging time of the measurement. Assuming that detection is limited by photon shot noise, it can be written that \cite{Shah}:
\begin{equation}
SNR = C \sqrt{\frac{P_l}{2 h \nu_l}}
\end{equation}
with $P_l$ the laser power at the output of the cell and $h \nu_l$ the thermal energy of a single photon. From measurements of $\Delta \nu$ and $C$ reported in Fig. \ref{fig:im3}, \ref{fig:Signal-OTS} and \ref{fig:C-OTS}, Fig. \ref{fig:Allan-OTS} shows the expected shot-noise limited clock frequency Allan deviation in CW regime versus the laser power. The best frequency stability result, at the level of 6 $\times$ 10$^{-13}$ at 1 s, is obtained for a laser power of about 300 $\mu$W. This result is about 3 times worse than the expected relative frequency stability of a clock using a buffer-gas filled Cs vapor cell of slightly smaller dimensions (10 mm length and 10 mm diameter) \cite{Liu:PRA:2013}.\\
Narrowing the linewidth
through the motion-induced Ramsey narrowing process could improve the frequency stability, but Breschi \textit{ et al.} \cite{Breschi:PRA:2010} have shown that the gain on the shot-noise limited frequency stability does not follow the gain on the resonance linewidth reduction. Actually, the same short-term relative frequency stability can be obtained with a large beam diameter at small laser intensity and a smaller beam diameter at higher laser intensity (i. e. higher clock signal).
With the Ramsey technique, the optimum clock short-term frequency stability is expected to be obtained when the free evolution time
$T_R$ equals the microwave coherence relaxation time $T_2$ \cite{Guerandel:IM:2007, Micalizio:Metrologia:2012, Liu:OE:2013}. In the present study, the free evolution time is limited to $T_2 \simeq 0.50$ ms, yielding a central fringe width of about 1 kHz.
The shot-noise limited frequency stability is given by \cite{Liu:OE:2013}:
\begin{equation}
\sigma_y (\tau) \approx \frac{\Delta \nu}{\nu_0} \frac{1}{C}\sqrt{\frac{h \nu_l}{P_l}}\sqrt{\frac{T_c}{t_m}} \tau^{-1/2},
\label{stabRam}
\end{equation}
with $T_c$ the length of an interrogation cycle, $t_m$ the length of the signal measurement. The stability limit in our case is about  $3.5 \times 10^{-12}$ at 1 s, worse than a CW interrogation, showing that a Ramsey interrogation is not the most appropriate in the case of a fast signal decay.\\

Eventually, from Fig. \ref{fig:Tshift-OTS1}, we note that for a cell temperature of 30$^{\circ}$C, the light shift slope is measured to be $-$0.029 Hz/$\mu$W, i.e $-$50.9 Hz/(mW/cm$^2$), which yields relatively to the clock frequency 5.5 $\times$ 10$^{-9}$ $/$ (mW/cm$^2$). This value is of the same order of magnitude than light-shift coefficients measured in Ref. \cite{Breschi:PRA:2010} in paraffin-coated cells. Moreover, we note in Fig. 6 that the temperature dependence of the clock transition in the Cs-OTS cell is roughly 24 Hz/$^{\circ}$C, which yields relatively to the clock frequency 2.7 $\times$ 10$^{-9}$/K. This sensitivity is about 10 times higher than the one measured (2.7 $\times$ 10$^{-10}$/K) in a laser-pumped paraffin-coated cell rubidium frequency standard \cite{Bandi:JAP:2012} and huge compared to typical sensitivities that can be achieved in optimized buffer-gas filled vapor cells \cite{Kozlova:PRA:2011, Vanier:JAP:1982, Miletic:EL:2010}. This high temperature sensitivity could be a serious drawback for the use of OTS-coatings towards the development of high-performance compact atomic clocks.

\section{Conclusions}
This paper reported the realization of a centimeter-scale Cs vapor cell wall-coated with octadecyltrichlorosilane (OTS) using a simple vapor phase deposition technique.
The coating properties were characterized by means of CPT spectroscopy of the Cs ground state hyperfine transition, in view of application to vapor cell atomic clocks. The presence of a narrow peak Lorentzian structure on the top of a Doppler-broadened broad structure is a relevant signature of the OTS anti-relaxation coating. Experimental results are given for a cell temperature of 35$^{\circ}$C, which  maximizes the CPT signal. Using clock frequency shift measurements, the adsorption energy of Cs atoms on OTS surface was measured to be $E_a= (0.42  \pm 0.03)$ eV, in agreement with the value reported in the literature for Cs-OTS-Pyrex \cite{Stephens:JAP:1994} by a completely different method. A clock frequency shift rate of 24.6 Hz/$^{\circ}$C was measured. The phase shift per collision with the OTS surface was estimated to $(-168 \pm 2)$ mrad, with a  shift rate of 7.4 mrad/$^{\circ}$C.\\ Measurements of hyperfine population lifetime ($T_1$) and microwave coherence lifetime ($T_2$) of about 1.6 and 0.5 ms were reported, corresponding to about 37 and 12 useful bounces, respectively. For comparison, in \cite{Seltzer:PRA:2007}, the reported number of bounces (for the lifetime $T_1$) for four OTS-coated K cells ranges from 20 to 2100. Using a Rb cell coated with an OTS monolayer, Ref. \cite{Yi:JAP:2008} reports 31 bounces (at 82$^{\circ}$C) and 41 bounces (at 102$^{\circ}$C) for Zeeman coherences, and 21 (at 102$^{\circ}$C) and 30 (at 170$^{\circ}$C) for hyperfine coherence. Ref. \cite{Straessle:APL:2014} reports for a Rb-OTS cell 11 bounces for the hyperfine coherence at a not clearly specified temperature, probably 60$^{\circ}$C. It appears that the number of useful bounces is smaller for hyperfine coherence than for Zeeman coherences, decreases at lower temperatures and decreases for atoms with higher polarizability. In that sense, the relatively low number of bounces (12) reported here for the hyperfine coherence of Cs at 35$^{\circ}$C is reasonable and give us confidence on the quality of the coating. \\
Motion-induced Ramsey narrowing \cite{Breschi:PRA:2010, Klein:PRA:2011} where atoms alternately spend bright and dark time intervals inside and outside the laser beam was observed in the Cs-OTS cell in continuous interaction regime by reducing the laser beam diameter. Raman-Ramsey CPT fringes were detected in the Cs-OTS cell using a temporal Ramsey-like pulsed interrogation scheme. Applications to atomic clocks were discussed. The best calculated shot-noise limited frequency stability was obtained for a CW interrogation. \\
As a matter of fact, the high Cs-OTS adsorption energy leads to a large atom-wall collision induced phase shift, a high temperature sensitivity of the clock transition frequency, $2.7\times10^{-9}/^{\circ}$C in fractional frequency, and a limited number (about 12) of bounces preserving the hyperfine coherence. We consider these aspects as serious drawbacks for the development of a high-performance Cs-OTS atomic clocks compared to the buffer gas technique. Nevertheless, OTS coatings could be remain of interest in microfabricated cells and miniature atomic clocks applications. Miniaturized cells usually work at higher temperature ($\sim$ 80 $-$ 100$^{\circ}$C) than centimeter-sized cells in order to increase the Cs density and the signal. As shown in Fig. \ref{fig:phase-T}, the module of the phase shift per collision decreases when the temperature increases. If the microcell operates at a temperature where the phase shift value is reduced or canceled, the broadening term  $\Delta \nu_{\Phi}$ can be reduced and consequently the number of useful bounces increased. On the other hand, the temperature-related shift will be certainly reduced but not canceled. The use of OTS coatings will be compliant with miniature atomic clocks applications if both the sensitivity of the clock frequency to cell temperature variations and cell temperature control are compatible with typical clock relative frequency stability performances objectives at the level of 10$^{-11}$ at 1 hour and 1 day averaging time.

\section*{Acknowledgments}\label{sec:Acknow}
We would like to thank P. Bonnay (LNE-SYRTE) for the Cs filling of Cs vapor cells, F. Peirera Dos Santos (LNE-SYRTE) and M. Hasegawa (EMPA) for helpful discussions. M. Abdel Hafiz PhD thesis is co-funded by the R\'egion Franche-Comt\'e and the LabeX FIRST-TF (Facilities for Innovation, Research, Services, Training in Time \& Frequency). V. Maurice PhD thesis is funded by the D\'el\'egation G\'en\'erale de l'Armement (DGA) and the R\'egion Franche-Comt\'e. This work is supported in part by the Agence Nationale de la Recherche and the DGA (ISIMAC project ANR-11-ASTR-0004).

\clearpage

\section*{List of Figure Captions}

Fig. 1. (Color online) Photograph of the OTS-coated cell.\\

Fig. 2. (Color online) Experimental setup used to detect CPT resonances in Cs vapor cells. DFB: Distributed-Feedback diode laser, MZ EOM: Mach-Zehnder electro-optic modulator, Michelson: Michelson delay-line and polarization orthogonalizer system, $V_{bias}$: dc bias voltage applied onto the EOM to reject the optical carrier, LO: local oscillator, FC: fiber collimator, PD: photodiode, LA1: lock-in amplifier, PC: personal computer. A microwave switch can be used to turn on and off the 4.596 GHz signal in order to make the atoms interact with a sequence of optical pulse trains.\\

Fig. 3. (Color online) CPT clock resonance in the Cs-OTS cell compared to the one detected in the evacuated Cs cell. The laser power incident on the cell is 100 $\mu$W. The cell temperature is 35$^{\circ}$C. Experimental data for the evacuated Cs cell are approximated by a Lorentzian fit function. The narrow structure of the resonance in the Cs-OTS cell (figure inset) is approximated by a Lorentzian fit function. The laser beam diameter is 15 mm. For the Cs-OTS cell, the slight asymetry of the CPT resonance is attributed to a minor thermal transient of the clock set-up table during the scan of the CPT resonance.\\

Fig. 4. Amplitude of the CPT signal (narrow structure) in the Cs-OTS cell versus the cell temperature. The laser power is 600 $\mu$W. The static magnetic field is 892 mG.\\

Fig. 5. (Color online) Linewidth of the CPT resonance in vapor cells versus the laser power for: evacuated Cs cell (squares), Cs-OTS cell narrow structure (circles), Cs-N$_2$-Ar cell (stars). Error bars are covered by data points. The cell temperature is 35$^{\circ}$C. Data of the evacuated Cs cell are fitted by the function $a+b\sqrt{P}+c P$ (dashed line, $P$ the laser power, $a$, $b$, and $c$ are constants). Experimental data in the Cs-OTS cell and the buffer-gas filled cell are fitted by a linear function. The figure inset is a zoom for Cs-OTS and Cs-N$_2$-Ar cells.\\

Fig. 6. (Color online) Logarithm of the frequency shift $|\delta \nu|$ versus the inverse of the cell temperature $1/T$ with $T$ in Kelvins. Experimental data are fitted by a linear function from which is extracted the value of $E_a$. In the main figure, error bars are covered by data points. The inset figure shows an example for a cell temperature of 30$^{\circ}$C of light-shift measurement used to extract the temperature frequency shift at null laser power. \\

Fig. 7. (Color online) Phase shift per wall collision as a function of the cell temperature. The solid line is a linear fit. Error bars are covered by data points.\\

Fig. 8. (Color online) Measurement of the $T_1$ relaxation time in the Cs-OTS cell. The total laser power incident on the cell is 500 $\mu$W. The actual power seen by the atoms  (power of one sideband) is about one half of the total laser power. The cell temperature is 35$^{\circ}$C. The static magnetic field is 900 mG. Squares: data, solid line: fitted exponential function.\\

Fig. 9. Measurement of the $T_1$ relaxation time in the Cs-OTS cell versus the total laser power incident on the cell. The actual power seen by the atoms is about one half of the total laser power. The cell temperature is 35$^{\circ}$C. The static magnetic field is 900 mG.\\

Fig. 10. (Color online) (a): CPT resonance in the Cs-OTS cell for a beam diameter of 8 mm. The laser power is 100 $\mu$W. (b) Linewidth of the CPT resonance narrow structure versus the laser intensity for different laser beam diameters (8, 10 and 12 mm). Error bars are covered by data points.\\

Fig. 11.  CPT-Ramsey fringe detected in a Cs-OTS cell. $\tau_p$ = 2 ms,
$T_R$ = 0.2 ms. The laser power is 650 $\mu$W. The cell temperature is 35$^{\circ}$C. The beam diameter covers the whole cell diameter.\\

Fig. 12. (Color online)  Amplitude of the central CPT-Ramsey fringe detected in a Cs-OTS cell versus the Ramsey time
$T_R$. The laser power is 1.4 mW. Experimental data are fitted by an exponential decay function of time constant $T_2$ = 0.46 ms.\\

Fig. 13. (a): CPT resonance signal, (b) CPT resonance contrast, (c) calculated shot-noise limited clock frequency Allan deviation versus the laser power incident on the cell. The cell temperature is 35$^{\circ}$C. The laser beam diameter equals the cell diameter. Error bars are covered by data points.\\


\clearpage
\section*{List of Figure Captions}

\clearpage

\clearpage

\clearpage

\clearpage
\clearpage
\clearpage

\clearpage
\clearpage
\clearpage
\clearpage

\clearpage
\clearpage

\end{document}